\documentclass{emulateapj}
\usepackage{apjfonts}
\usepackage{xspace}
\usepackage{amsmath}
\usepackage{mathtools}

\usepackage[usenames,dvipsnames]{xcolor}

\usepackage{etoolbox}
\makeatletter
\patchcmd{\NAT@citex}
  {\@citea\NAT@hyper@{\NAT@nmfmt{\NAT@nm}\NAT@date}}
  {\@citea\NAT@nmfmt{\NAT@nm}\NAT@hyper@{\NAT@date}}
  {}
  {}
\patchcmd{\NAT@citex}
  {\@citea\NAT@hyper@{%
     \NAT@nmfmt{\NAT@nm}%
     \hyper@natlinkbreak{\NAT@aysep\NAT@spacechar}{\@citeb\@extra@b@citeb}%
     \NAT@date}}
  {\@citea\NAT@nmfmt{\NAT@nm}%
   \NAT@aysep\NAT@spacechar%
   \NAT@hyper@{\NAT@date}}
  {}
  {}
\patchcmd{\NAT@citex}
  {\@citea\NAT@hyper@{%
     \NAT@nmfmt{\NAT@nm}%
     \hyper@natlinkbreak{\NAT@spacechar\NAT@@open\if*#1*\else#1\NAT@spacechar\fi}%
       {\@citeb\@extra@b@citeb}%
     \NAT@date}}
  {\@citea\NAT@nmfmt{\NAT@nm}%
   \NAT@spacechar\NAT@@open\if*#1*\else#1\NAT@spacechar\fi%
   \NAT@hyper@{\NAT@date}}
  {}
  {}
\makeatother

\begin{document}

\title{Weighing Galaxy Clusters with Gas. II. \\ On the Origin of Hydrostatic Mass Bias in $\Lambda$CDM Galaxy Clusters}
\shorttitle{Origin of Hydrostatic Mass Bias}
\shortauthors{Nelson et al.}
\slugcomment{{\sc Accepted to ApJ:} January 6, 2014}

\author{Kaylea Nelson\altaffilmark{1}}
\author{Erwin T. Lau\altaffilmark{2,3}}
\author{Daisuke Nagai\altaffilmark{1,2,3}}
\author{Douglas H. Rudd\altaffilmark{2,3,4,5}}
\author{Liang Yu\altaffilmark{1,2}}

\affil{ 
$^1$Department of Astronomy, Yale University, New Haven, CT 06520, U.S.A.; \href{mailto:kaylea.nelson@yale.edu}{kaylea.nelson@yale.edu} \\
$^2$Department of Physics, Yale University, New Haven, CT 06520, U.S.A.\\
$^3$Yale Center for Astronomy \& Astrophysics, Yale University, New Haven, CT 06520, U.S.A. \\
$^4$Kavli Institute for Cosmological Physics, University of Chicago, Chicago, IL 60637, U.S.A. \\
$^5$Research Computing Center, University of Chicago, Chicago, IL 60637, U.S.A.
}


\begin{abstract}
The use of galaxy clusters as cosmological probes hinges on our ability to measure their masses accurately and with high precision. Hydrostatic mass is one of the most common methods for estimating the masses of individual galaxy clusters, which  suffer from biases due to departures from hydrostatic equilibrium. Using a large, mass-limited sample of massive galaxy clusters from a high-resolution hydrodynamical cosmological simulation, in this work we show that in addition to turbulent and bulk gas velocities, acceleration of gas introduces biases in the hydrostatic mass estimate of galaxy clusters. In unrelaxed clusters, the acceleration bias is comparable to the bias due to non-thermal pressure associated with merger-induced turbulent and bulk gas motions. In relaxed clusters, the mean mass bias due to acceleration is small ($\lesssim 3$\%), but the scatter in the mass bias can be reduced by accounting for gas acceleration. Additionally, this acceleration bias is greater in the outskirts of higher redshift clusters where mergers are more frequent and clusters are accreting more rapidly. Since gas acceleration cannot be observed directly, it introduces an {\em irreducible} bias for hydrostatic mass estimates. This acceleration bias places limits on how well we can recover cluster masses from future X-ray and microwave observations.  We discuss implications for cluster mass estimates based on X-ray, Sunyaev-Zeldovich effect, and gravitational lensing observations and their impact on cluster cosmology. 
\end{abstract}

\section{Introduction}

Galaxy clusters are the largest gravitationally bound objects in the universe and as such are excellent probes of growth of structure and dark energy \citep[see e.g.,][for review]{allen_etal11}. Their abundance depends sensitively on cosmology and has to date provided meaningful cosmological constraints \citep[e.g.][]{allen_etal08, vikhlinin_etal09, planck_XX13} that make clusters competitive cosmological tools.  However, in order to harness the full potential of clusters to further our understanding of cosmology in upcoming cluster surveys (e.g., {\em eROSITA}), we must be able to make accurate measurements of the cluster mass function and its evolution in order to reduce systematic uncertainties in cosmological parameters.

There are a number of techniques for observationally determining the masses of clusters. Currently one of the most accurate mass measurements of individual clusters is through X-ray observations. Assuming the intracluster medium (ICM) is in hydrostatic equilibrium (HSE) with the cluster gravitational potential, one can estimate cluster mass with measurements of the density and temperature profiles of the ICM.  However, the application of the hydrostatic mass estimate is based on the assumption that the ICM is both spherically symmetric and in hydrostatic equilibrium. In the hierarchal structure formation model, galaxy clusters, having recently formed, are dynamically active systems and are not in exact hydrostatic equilibrium. This leads to a bias in hydrostatic mass estimate. Both observations (comparisons of lensing to X-ray) and hydrodynamical simulations have found that the hydrostatic mass underestimates the true cluster mass by 5\%--30\% \citep[e.g.,][]{rasia_etal06,nagai_etal07b, jeltema_etal08, piffaretti_etal08, zhang_etal10, meneghetti_etal10,becker_kravtsov11,rasia_etal12, mahdavi_etal13}. Recent works suggest that this bias stems from neglecting non-thermal pressure support, which are mostly provided by bulk and turbulent gas flows generated primarily by mergers and accretion during cluster formation \citep{lau_etal09, vazza_etal09, nelson_etal12}.  By including the pressure support from such gas motions, it is possible to at least partially recover the true mass of the system, depending on the dynamical state of the cluster \citep{rasia_etal04, fang_etal09, lau_etal09, nelson_etal12}.

Previous attempts to correct the hydrostatic mass bias by accounting for gas motions have operated under the assumption of steady state where net gas velocity is constant with time. A recent work by \citet{suto_etal13} relaxed this assumption and showed that the mass contribution from gas accelerations can be non-negligible. In \citet[][hereafter Paper I]{lau_etal13} we showed that the inclusion of the additional term due to gas acceleration becomes particularly important in the outskirts of relaxed clusters.  Observationally, it is very difficult to measure gas acceleration, and therefore numerical simulations are useful for characterizing gas acceleration and its effect on the hydrostatic mass. However, previous theoretical works on gas acceleration were limited to only a few clusters. Since the gas acceleration is likely sensitive to the rate at which the cluster is accreting materials in the cluster outskirts, it is important to characterize the gas acceleration for a wide range of mass accretion histories using a large cosmologically representative sample of galaxy clusters. 

In this work we present a large high-resolution hydrodynamical cosmological simulation to characterize the hydrostatic mass bias for a wide range of masses, redshifts, and cluster dynamical states. Our large uniform mass-limited sample contains 62 clusters with $M_{500}\footnotemark\ \geq 3\times 10^{14}h^{-1}M_{\odot}$, which is cosmologically representative in terms of cluster dynamical state, critical for examining the effects of mergers and accretion on hydrostatic mass bias and on the contribution of gas acceleration. Compared to similar previous works with  comparable numbers of clusters \citep[e.g.,][]{jeltema_etal08, piffaretti_etal08}, our simulated cluster sample has improved by over an order of magnitude in both mass and spatial resolution and better resolves the gas flows responsible for the hydrostatic mass bias. We quantify the dynamical states of our systems by constructing merger trees and tracking the most massive progenitors of the $z=0$ clusters. We then compute the fractional change in their mass from $z = 0.5$, a quantity that is sensitive to the mass assembly histories of galaxy clusters. 

In Paper I, we presented two methods of reconstructing cluster masses using gas information, the ``summation'' method and the ``averaging'' method that are shown to be mathematically equivalent. In this work we adopt the averaging method for computing the hydrostatic mass bias, as it more closely resembles observational procedures which measure spatially averaged quantities. We show that in unrelaxed clusters the mass bias due to gas acceleration is comparable to the bias due to non-thermal pressure associated with merger-induced turbulent and bulk gas motions. In relaxed clusters, the bias due to gas acceleration is small  ($\lesssim 3\%$) on average, but the scatter in the mass bias can be reduced by accounting for acceleration.  Since it is not directly observable, gas acceleration introduces an \emph{irreducible} bias in the hydrostatic mass estimates of galaxy clusters.

This paper is organized as follows. In Section 2 we summarize the method of cluster mass reconstruction, presented in full in Paper I. In Section 3, we describe our simulated cluster sample. We present our analysis of the mass reconstructions and the acceleration term in Section 4.  Our results and their implications are summarized in Section 5.

\footnotetext[1]{Throughout this paper, we refer to total mass and ICM properties within radii which correspond to fixed overdensities $\Delta$ relative to the critical density at that redshift, such that $M_{\Delta} (r_{\Delta}) = \Delta(4/3)\pi r^3_{\Delta}\rho_c(z)$.  All masses stated herein are calculated within the true value of $r_{\Delta}$ as measured directly from the simulations.} 

\section{Theoretical framework for mass reconstruction}
\label{sec:theory}

Here we provide a brief overview of the method for computing cluster mass using gas information presented in Paper I. Using Gauss's law for the gravitational field, the total gravitational mass enclosed within volume $V$ with surface $\partial V$ is 
\begin{equation}
M = \frac{1}{4\pi G}\int_{\partial V} \nabla \Phi \cdot d{\bf S} 
\label{eqn:gauss}\\
\end{equation}
where $M$ is the enclosed mass and $\Phi$ is the gravitational potential. The mass inside this surface can obtained when the potential gradient $\nabla \Phi$ is known at every position on the imaginary surface with surface element $d{\bf S}$. 

In hydrodynamical simulations without physical viscosity, each gas element follows the Euler equation:
\begin{equation}
\frac{\partial u^{i}}{\partial t} + u^j\frac{\partial u^{i} }{\partial x^j} = -\frac{1}{\rho}\frac{\partial P}{\partial x_i} -\frac{\partial \Phi}{\partial x_i}.
\label{eqn:euler}
\end{equation}
Combined with Gauss's Law (Equation~\ref{eqn:gauss}), the mass is given by 
\begin{equation}
M= -\frac{1}{4\pi G}\int_{\partial V} \left(\frac{\partial u^i}{\partial t}+u^j\frac{\partial u^i}{\partial x^j}+\frac{1}{\rho}\frac{\partial P}{\partial x_i}\right) dS_i .
\end{equation}
In practice, we often have access to only averaged quantities. Spatially averaging the above over the spherical surface with radius $r$ leads to
\begin{align}
M(<r) = M_{\rm tot}(<r) &= M_{\rm therm} + M_{\rm rand} + M_{\rm rot} \nonumber \\ 
&+ M_{\rm cross}+ M_{\rm stream} +M_{\rm accel},
\label{eqn:mass_recon}
\end{align}
where
\begin{align}
&M_{\rm therm} = \frac{-r^2}{G\langle\rho\rangle}\frac{\partial \langle P \rangle}{\partial r}, \label{eqn:jeans_therm}\\
&M_{\rm rand} = \frac{-r^2}{G\langle\rho\rangle}\frac{\partial \langle \rho \rangle \sigma_{\rho,rr}^{2}}{\partial r} - \frac{r}{G}\left(2\sigma_{\rho,rr}^2-\sigma_{\rho,\theta\theta}^2-\sigma_{\rho,\phi\phi}^2\right), \label{eqn:jeans_rand}\\ 
&M_{\rm rot} = \frac{r}{G}\left(\langle u_\theta\rangle_\rho^2+\langle u_\phi\rangle_\rho^2\right),  \label{eqn:jeans_rot}\\
&M_{\rm stream} = \frac{-r^2}{G}\left(\langle u_r\rangle_\rho\frac{\partial \langle u_r\rangle_\rho }{\partial r} + \frac{\langle u_\theta\rangle_\rho}{r}\frac{\partial \langle u_r\rangle_\rho }{\partial \theta} +\frac{\langle u_\phi\rangle_\rho}{r\sin \theta}\frac{\partial \langle u_r\rangle_\rho }{\partial \phi}\right),  \label{eqn:jeans_stream}\\
&M_{\rm cross} = \frac{-r^2}{G\langle\rho\rangle}\left(\frac{1}{r}\frac{\partial \langle \rho \rangle \sigma_{\rho,r\theta}^{2}}{\partial \theta} + \frac{1}{r\sin\theta}\frac{\partial \langle \rho \rangle \sigma_{\rho,r\phi}^{2}}{\partial \phi}\right) \nonumber \\
&\qquad \quad {}-\frac{r}{G}\left( \sigma_{\rho,r\theta}^2\cot\theta \right),  \label{eqn:jeans_cross}\\
&M_{\rm accel} =  \frac{-r^2}{G}\frac{\partial \langle u_{r}\rangle_\rho}{\partial t}. \label{eqn:jeans_accel}
\end{align}
The physical meaning of the terms are as follows: 
$M_{\rm therm}$ is the term representing the support against gravity from the averaged thermal pressure of the gas;
$M_{\rm rand}$ is the support from the random motions of gas in both the radial and tangential directions;
$M_{\rm rot}$ is the rotational support due to {\em mean} tangential motions of gas;
$M_{\rm stream}$ comes from spatial variations of the {\em mean} radial streaming gas velocities; 
$M_{\rm cross}$ arises from the off-diagonal components of the velocity dispersion tensor, which are non-zero if the radial and tangential components of the random motions are correlated; and $M_{\rm accel}$ is the support due to to temporal variations of the {\em mean} radial gas velocities at a fixed radius, which is negative (positive) for net gas accelerating (decelerating) away from the cluster center.

In this work, we are primarily interested in how the hydrostatic mass ($M_{\rm therm}$), total recovered mass ($M_{\rm tot}$), and the effective mass term from gas acceleration ($M_{\rm accel}$) depend on the dynamical state of the cluster. 

\section{Simulations}
\label{sec:data}

\subsection{Hydrodynamical Simulations of Galaxy Clusters}

In this work we analyze a high-resolution cosmological simulation of 62 galaxy clusters in a flat $\Lambda$CDM model with WMAP five-year ({\em WMAP5}) cosmological parameters: $\Omega_M = 1 - \Omega_{\Lambda} = 0.27$, $\Omega_b = 0.0469$, $h = 0.7$ and $\sigma_8 = 0.82$, where the Hubble constant is defined as $100h$~km~s$^{-1}$~Mpc$^{-1}$ and $\sigma_8$ is the mass variance within spheres of radius 8$h^{-1}$ Mpc. The simulation is performed using the Adaptive Refinement Tree (ART) $N$-body+gas-dynamics code \citep{kra99, kra02, rudd_etal08}, which is an Eulerian code that uses adaptive refinement in space and time, and non-adaptive refinement in mass \citep{klypin_etal01} to achieve the dynamic ranges to resolve the cores of halos formed in self-consistent cosmological simulations. The simulation volume has a comoving box length of $500\,h^{-1}$~Mpc, resolved using a uniform $512^3$ grid and 8 levels of mesh refinement, implying a maximum comoving spatial resolution of $3.8\,h^{-1}$~kpc.  We selected clusters with $M_{500} \geq 3\times10^{14} h^{-1}M_{\odot}$ and performed a simulation where only the regions surrounding the selected clusters are resolved. The resulting simulation has effective mass resolution of $2048^{3}$ surrounding the selected clusters, allowing a corresponding mass resolution of $1.09 \times 10^9\, h^{-1}M_{\odot}$. The current simulation only models gravitational physics and non-radiative hydrodynamics. As shown in \citet{lau_etal09}, the exclusion of cooling and star formation have negligible effect (less than a few percent) on the total contribution of gas motions to the hydrostatic mass bias outside cluster cores. 

\subsection{Cluster Finder}

Galaxy clusters are identified in the simulation using a variant of the method 
described in \citet{tinker_etal08}. Potential clusters are identified as peaks 
in the dark matter distribution, found by constructing a local density estimate 
at the position of each dark matter particle using an SPH kernel and the 24 
nearest neighboring particles. For each potential cluster center, we grow a sphere at the
location of the particle with highest density enclosing an overdensity $500\rho_c(z)$ 
(where $\rho_c(z)$ is the critical density of the universe at redshift $z$),
including all matter components in the simulation.  We then apply an iterative
procedure to refine the cluster center by alternately reducing the current radius
by 5\% and shifting to the center of mass within that sphere.  This iteration
avoids mistakenly centering the cluster at the position of a massive substructure 
with higher central dark matter density.  We consider the center to be converged when 
it has moved by less than 5 times the minimum cell size or less than 
$10^{-4}$ of the current radius.

We recompute $r_{500}$ at the new cluster center and eliminate all other particles
within that radius as potential centers.  The cluster is discarded if its center lies 
within the $r_{500}$ of a previously identified cluster or if its $M_{500} < 10^{11} h^{-1}M_{\odot}$.  
The entire procedure is repeated for the next most dense dark matter particle until all
potential centers have been associated with a cluster or eliminated. This is a 
computationally efficient mechanism for identifying isolated clusters in a simulation 
containing both $N$-body and mesh mass components.

\begin{figure}[t]
\begin{center}
\includegraphics[scale=0.85]{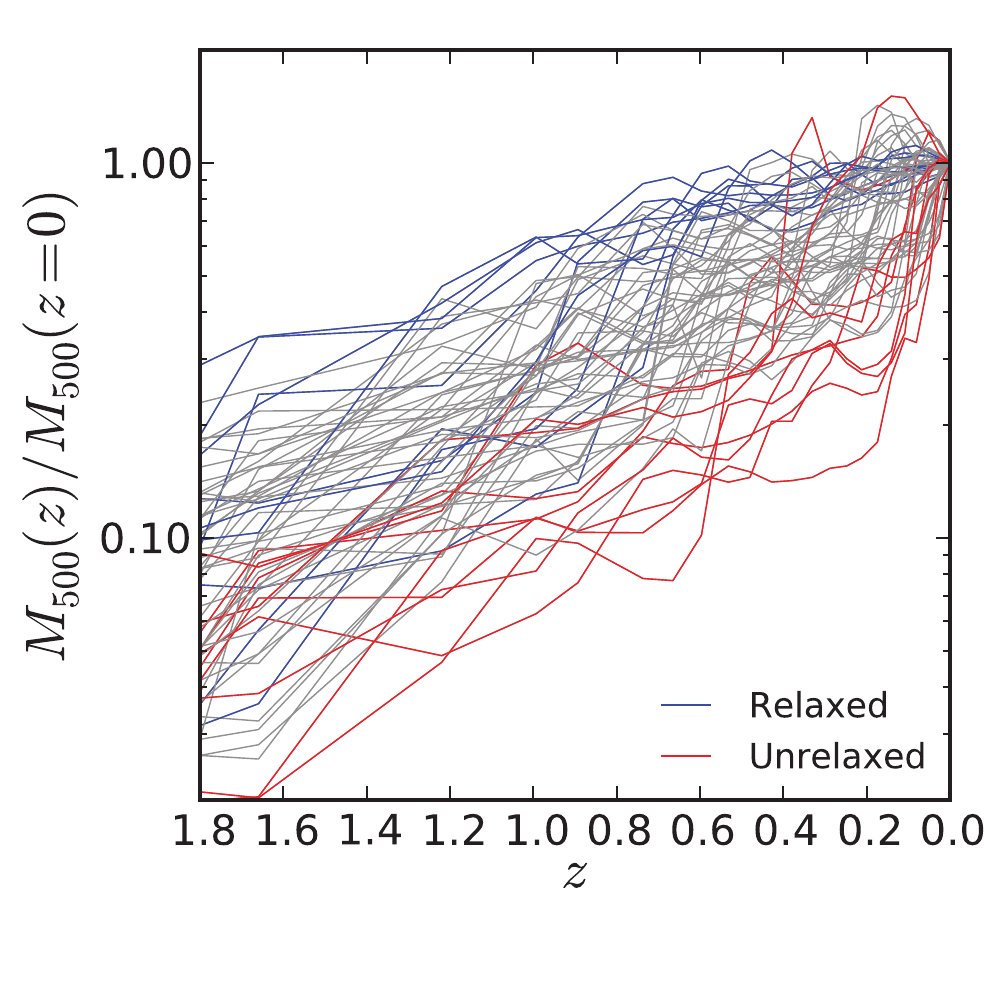}\vspace{-0.8cm}
\caption
{Mass accretion histories for relaxed (blue) and unrelaxed (red) clusters. Clusters are classified as relaxed if their mass accretion since $z$ = 0.5 is in the lowest 15\% of the sample, and unrelaxed if their mass accretion is in the highest 15\%.}
\label{fig:accretion_history}
\end{center}
\end{figure}

\subsection{Dynamical State}
\label{dynamicalstate}

We use each cluster's mass accretion history to identify its dynamical state at the present epoch. We identify and track the most massive progenitor of the $z=0$ clusters by iteratively following the dark matter particles in the clusters at each timestep to $z=0.5$. In the event of a merger, we follow the accretion history of the more massive progenitor.  We calculate the fractional increase in each progenitor's mass between the two epochs, $\Delta M_{500} \equiv M_{500}(z=0)/M_{500}(z=0.5)$. Clusters are then classified as relaxed or unrelaxed if their fractional mass growth are in the lowest or highest 15\% of the sample, respectively. Figure~\ref{fig:accretion_history} shows the mass accretion histories of the 62 clusters in our simulation sample. 

This method of characterizing cluster dynamical state is sensitive to the overall mass accretion history of each cluster, rather than recent merger history. To test the robustness of our ``relaxedness" selection criterion, we have compared our results to the time since last major merger method presented in \citet[][hereafter N12]{nelson_etal12} as well as varying the definition of relaxed cluster from bottom 10\% to 50\% of  $\Delta M_{500}$. A detailed examination of this comparison can be found in the \hyperref[sec:appendix]{Appendix}. We find that our results are insensitive to our choice of method. In addition, we choose to define our subsamples as the lowest or highest 15\% of the sample as this percentage balances a statistically significant sample size of clusters while maximally reducing the contamination by clusters with intermediate dynamical states.

\subsection{Method}

To compute each mass term in Section~\ref{sec:theory}, we work in the spherical coordinate system $(r,\theta,\phi)$, and divide the analysis region into 99 spherical bins spaced logarithmically from $10\,h^{-1}{\rm kpc}$ to $10 \,h^{-1}{\rm Mpc}$ in the radial direction from the cluster center, defined as the position with the maximum gas binding energy. Each spherical bin is further subdivided into 60 and 120 uniform angular bins in the $\theta$ and $\phi$ directions, respectively. We choose the rest frame of the system to be the velocity of the center of mass of the cluster interior to each radial bin, and rotate the coordinate system for each radial bin such that the $z$-axis aligns with the axis of the total gas angular momentum of that bin. 

We compute gas density-weighted gas velocities, volume-weighted density and volume-weighted pressure averaged over the hydro cells residing in each angular bin. 
We remove large gas substructures that may bias the global gas pressure and velocity gradients by applying the clump exclusion method presented in \citet{zhuravleva_etal13}. In addition, we smooth each mass term by applying the Savitzky-Golay filter used in \citet{lau_etal09}. Finally, the true mass $M_{\rm true}$ is measured directly. 
The velocity and pressure derivatives are computed by differencing neighboring angular bins. We then compute each mass term in Equations~(\ref{eqn:jeans_therm})~--~(\ref{eqn:jeans_stream}) by averaging values of the angular bins over the radial bin. The acceleration term $M_{\rm accel}$ is computed by taking the difference of radial velocity in the same radial bin between two consecutive timesteps ($\sim 0.04$~Gyr).

\begin{figure*}[htbp]
\begin{center}
\includegraphics[trim=0cm 1.1cm 4.8cm 0cm, scale=0.55,clip=true]{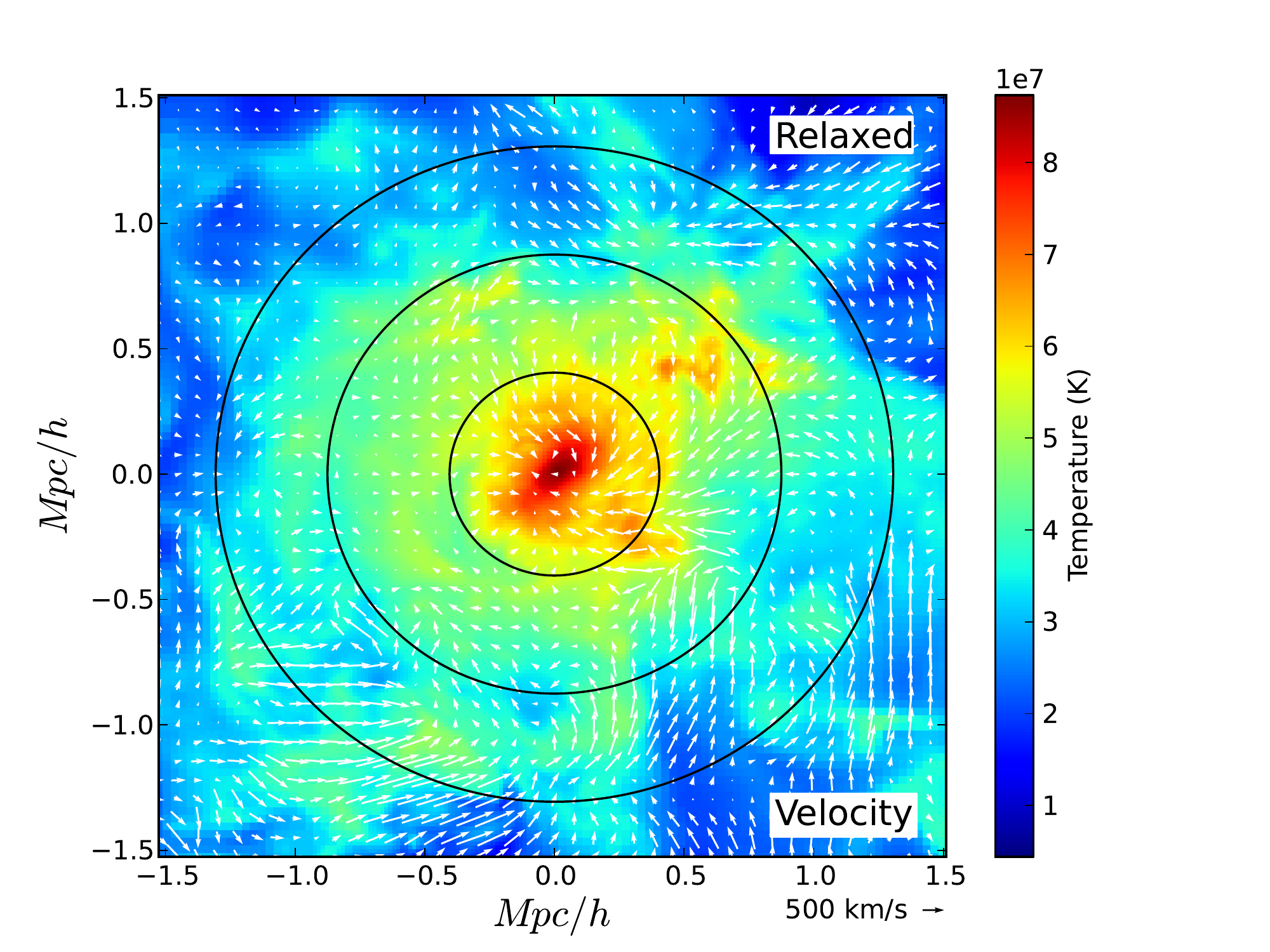}\includegraphics[trim=2cm 1.1cm 0cm 0cm, scale=0.55,clip=true]{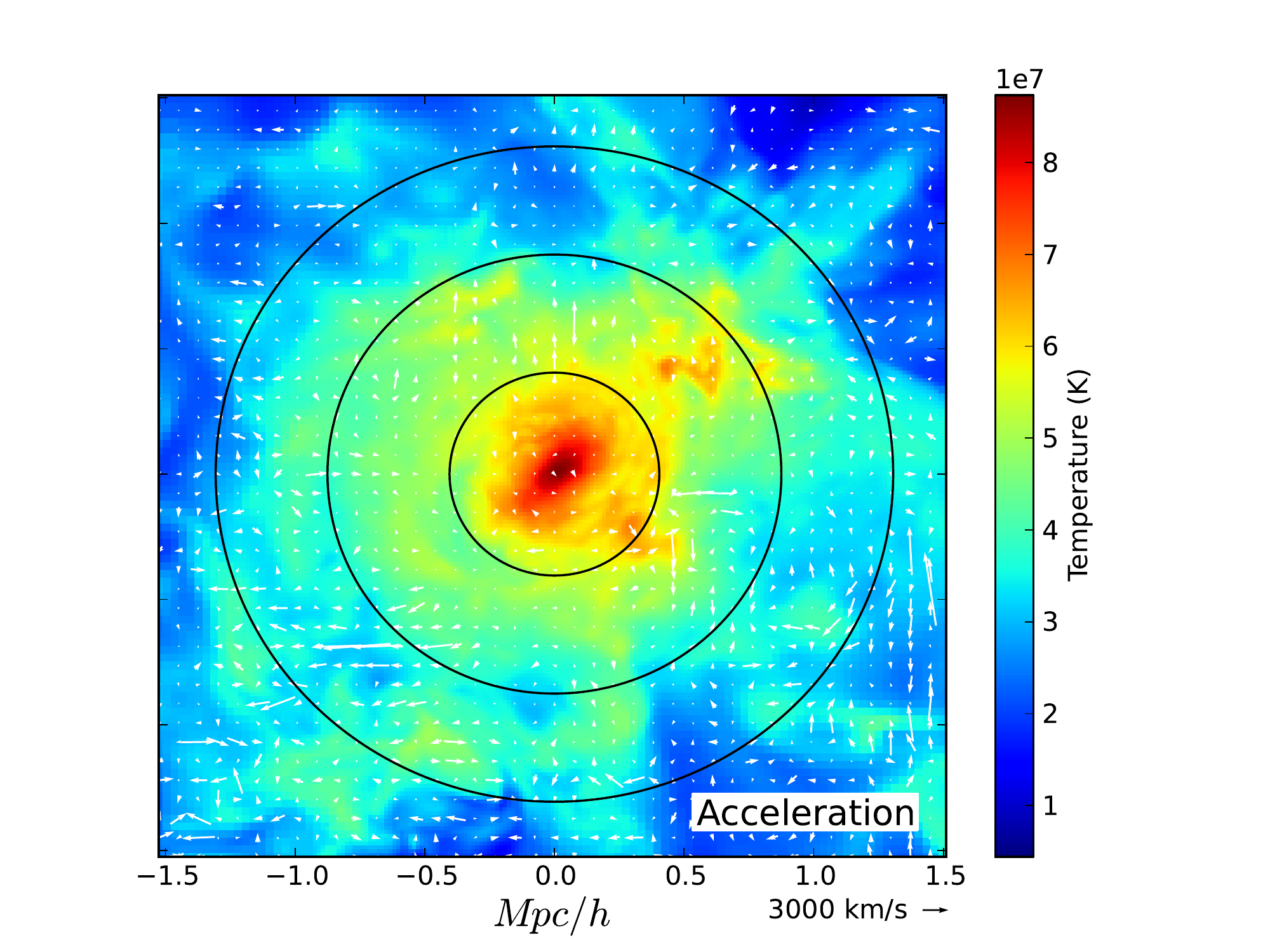}\vspace{-0.2cm}
\includegraphics[trim=0cm 0cm 4.8cm 0.8cm, scale=0.55,clip=true]{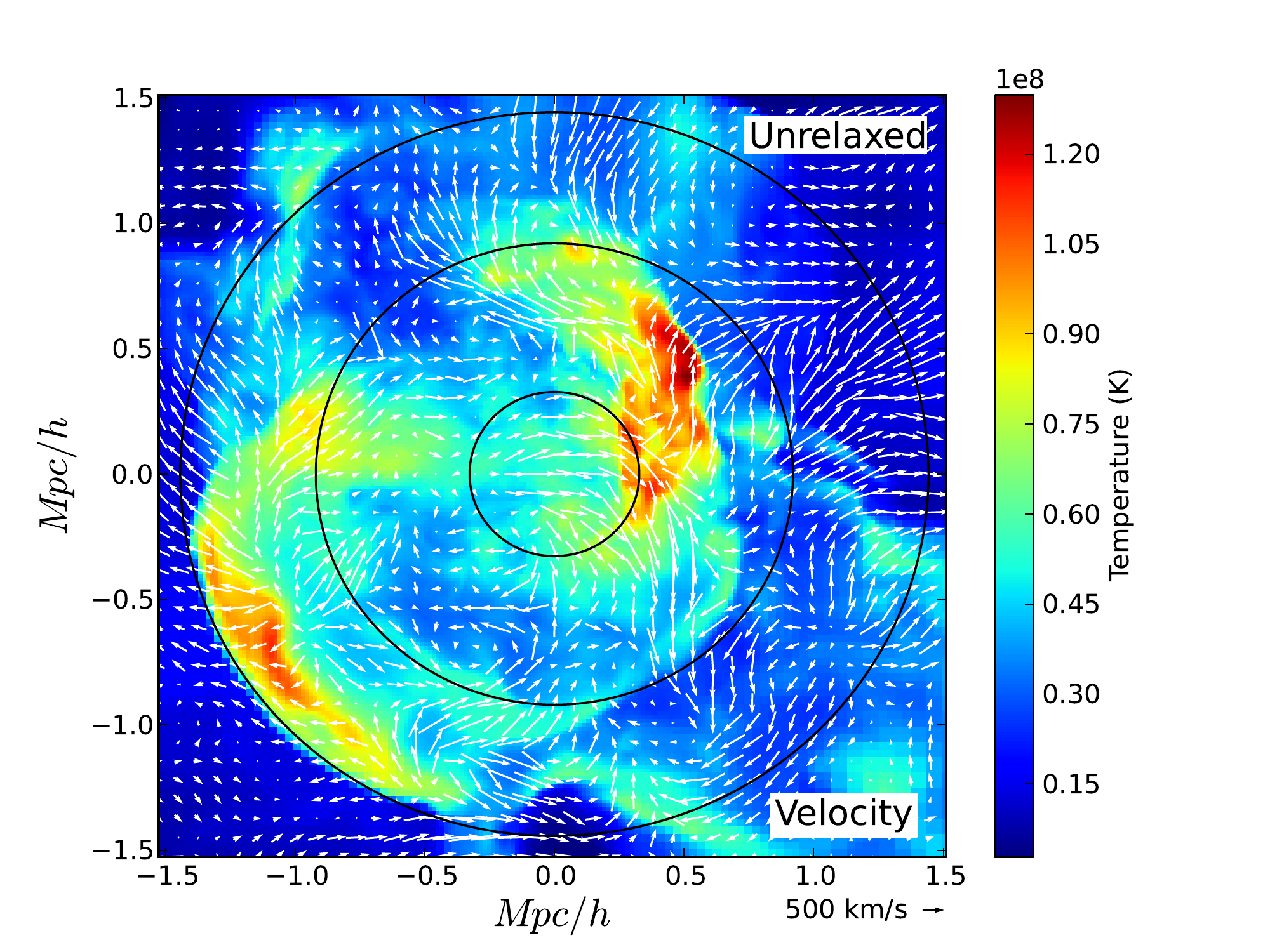}\includegraphics[trim=2cm 0cm 0cm 0.8cm, scale=0.55,clip=true]{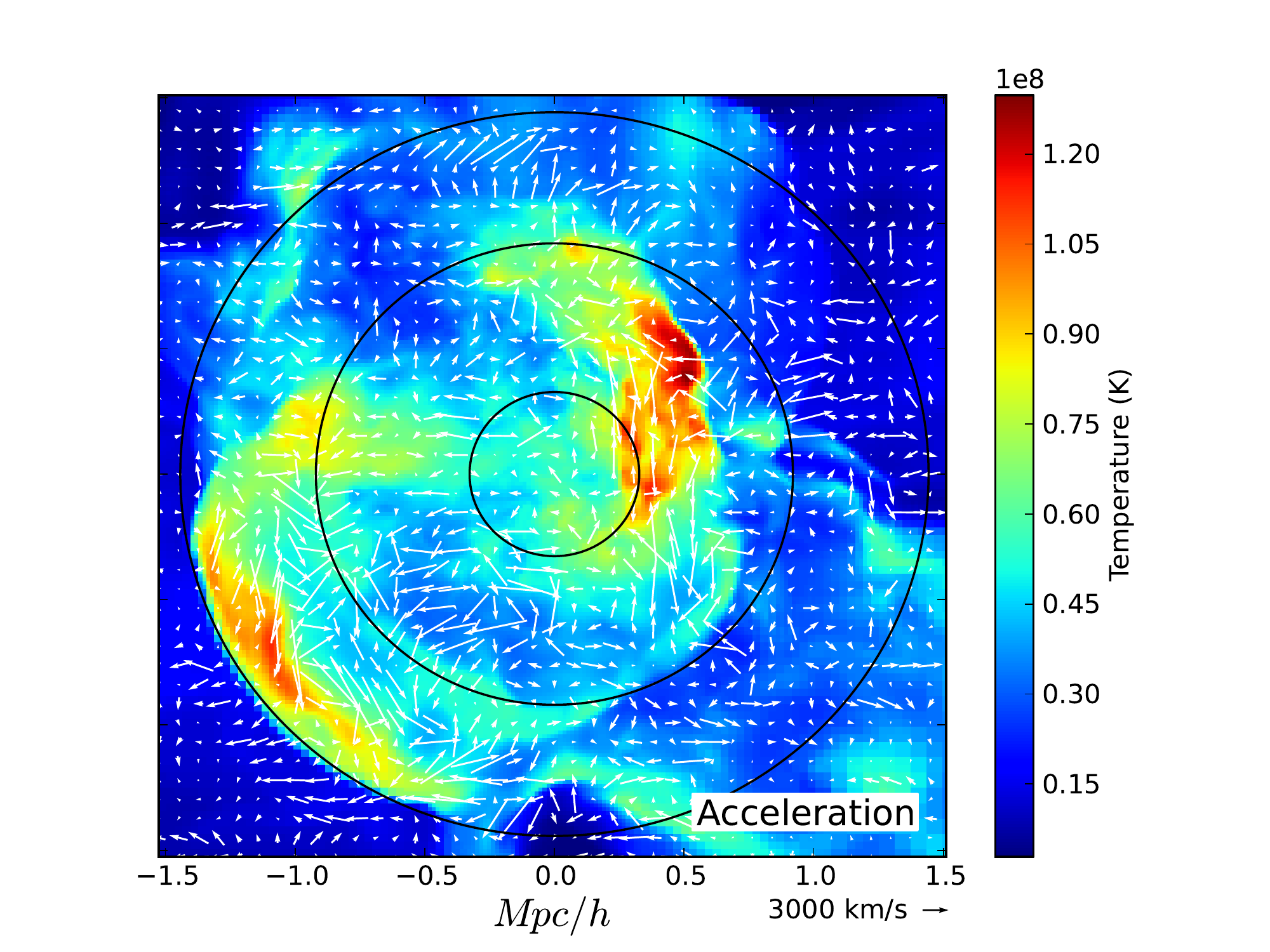}
\caption
{Projected mass-weighted temperature map of a relaxed ({\em top}) cluster and an unrelaxed ({\em bottom}) cluster with the velocity ({\em left}) and acceleration ({\em right}) vector fields overlaid. The black circles denote $r_{2500}$, $r_{500}$ and $r_{200}$ of the clusters from inside to outside. Both the maps and vector fields are mass weighted along a 200 kpc/h deep slice centered on their respective cluster centers.}
\label{fig:unrelaxed_example}
\end{center}
\end{figure*}

\section{Results}
\label{sec:results}

\begin{figure*}[htbp]
\begin{center}

\includegraphics[scale=0.63]{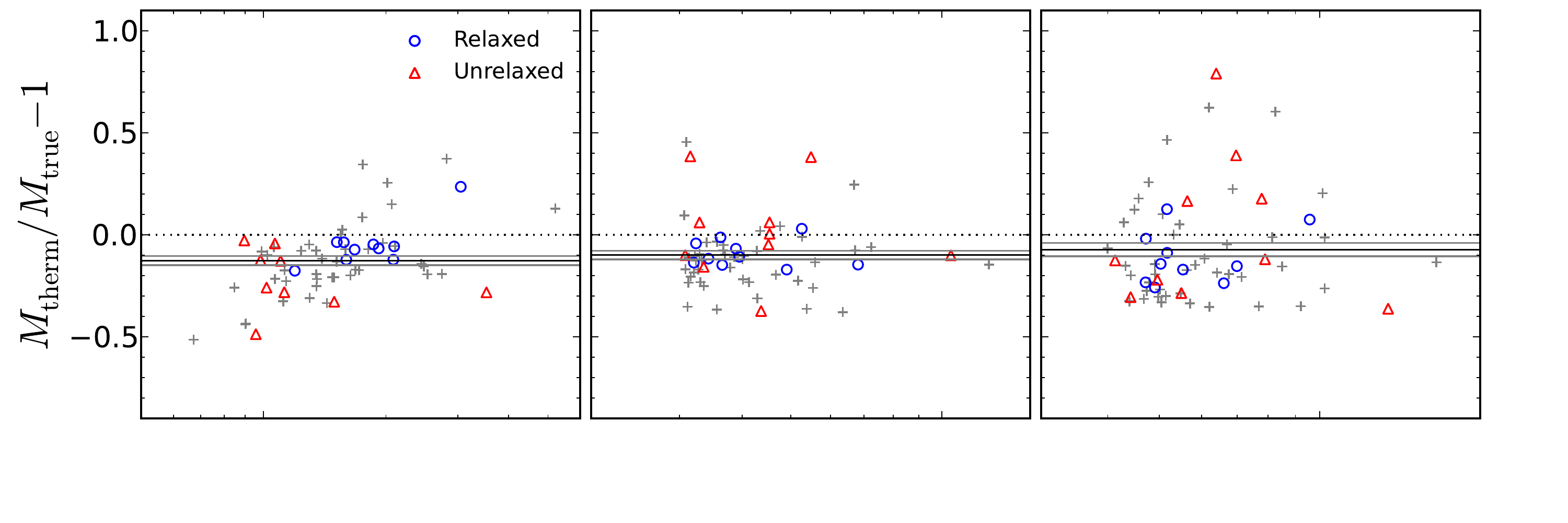}\vspace{-1.29cm}
\includegraphics[scale=0.63]{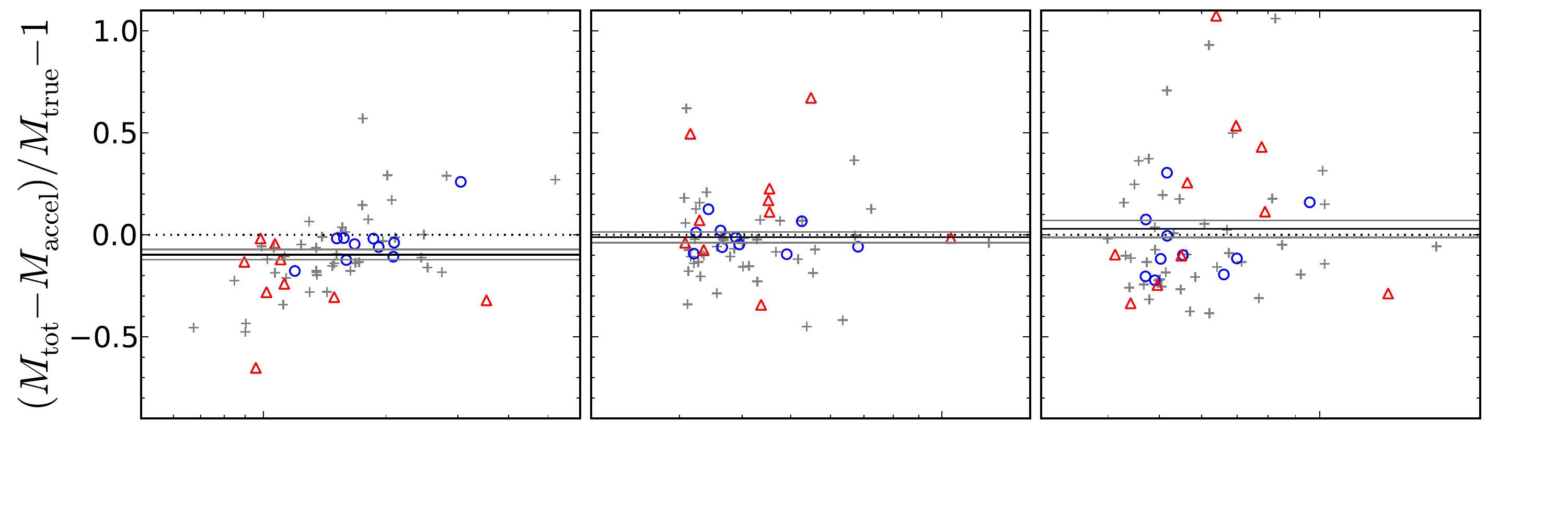}\vspace{-1.29cm}
\includegraphics[scale=0.63]{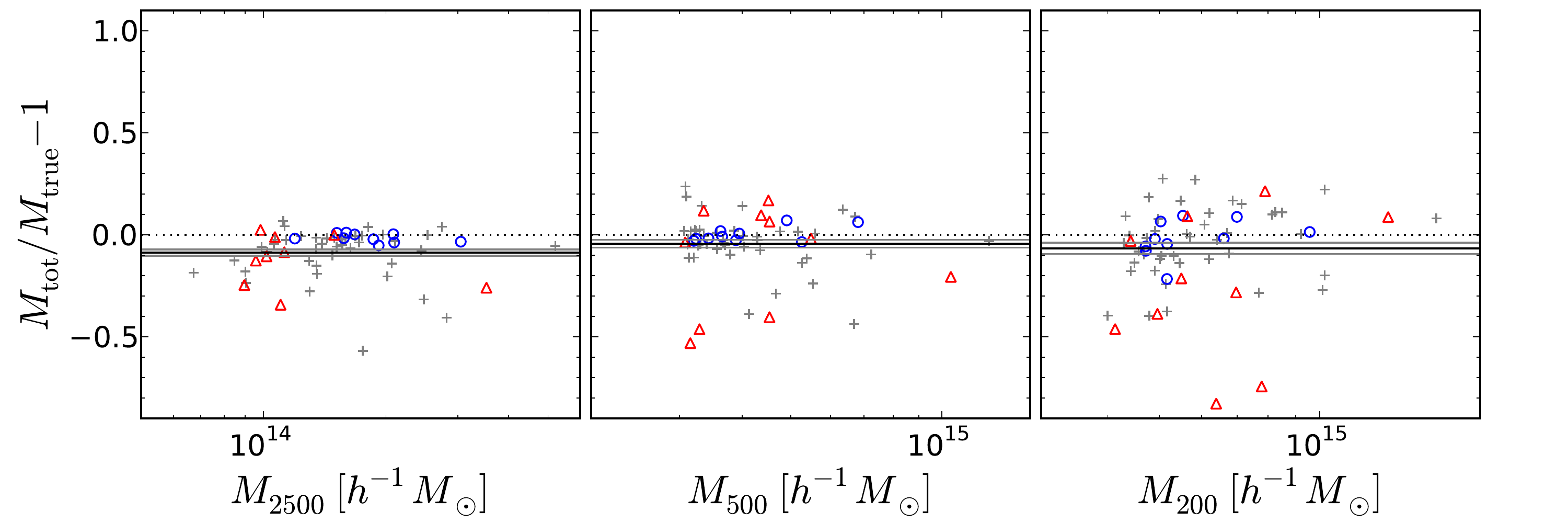}
\caption
{Comparison of the deviation of ({\em top to bottom}) $M_{\rm therm}$, $M_{\rm tot}$-$M_{\rm accel}$, and $M_{\rm tot}$ from the true mass for relaxed (blue circles) and unrelaxed (red triangles) clusters for three radii, $r_{2500}$, $r_{500}$ and $r_{200}$ ({\em left to right}). The remainder of the sample is marked by grey crosses.  The mean biases are denoted by black lines with the error (1$\sigma$) marked on either side by grey lines. Bias = 0 is depicted by the dotted lines. }
\label{fig:mcompare}
\end{center}
\end{figure*}

\begin{figure*}[htbp]
\begin{center}
\includegraphics[scale=0.63]{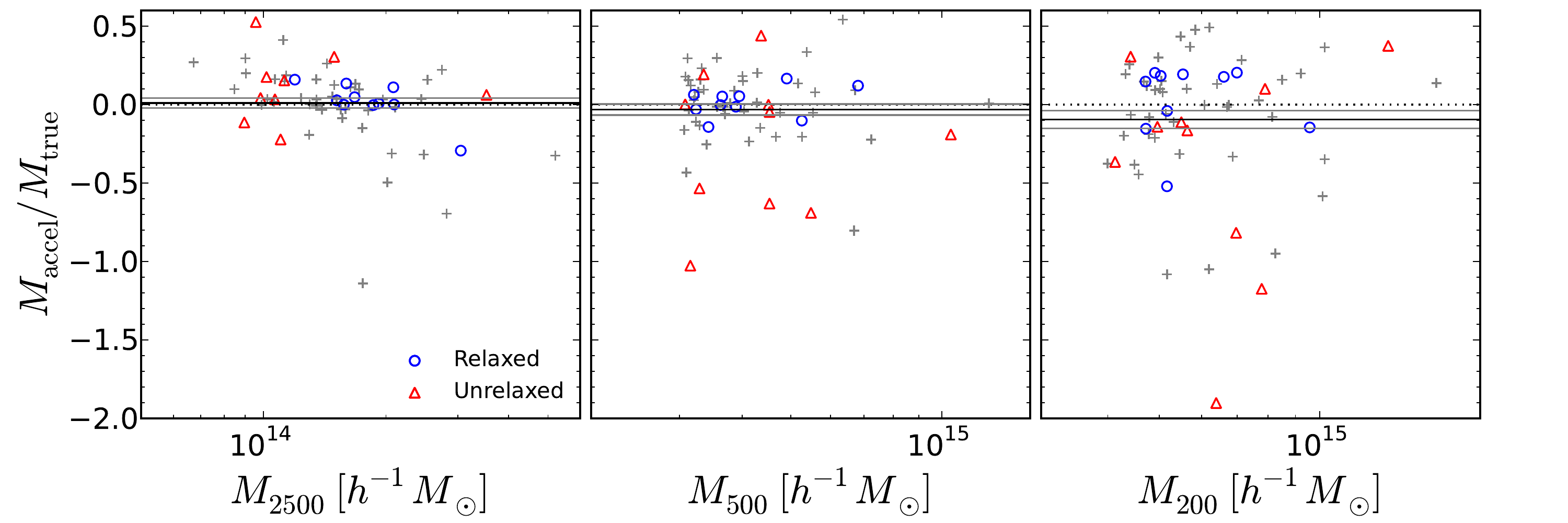}
\caption
{Comparison of fractional contribution of $M_{\rm accel}$ for relaxed (blue circles) and unrelaxed (red triangles) clusters for three radii, $r_{2500}$, $r_{500}$ and $r_{200}$ ({\em left to right}). The remainder of the sample is marked by grey crosses.  Mean $M_{\rm accel}$/$M_{\rm true}$ is denoted with a black line with the error (1$\sigma$) denoted on either side by grey lines. $M_{\rm accel}$/$M_{\rm true}$ = 0.0 is depicted by the dotted line.}
\label{fig:maccel}
\end{center}
\end{figure*}

First we show projected mass-weighted temperature maps of a relaxed cluster ({\em top}) and unrelaxed cluster ({\em bottom}) in  Figure~\ref{fig:unrelaxed_example}. To illustrate the complex velocity and acceleration structure in the ICM we have overplotted the velocity and acceleration vectors on the left and right hand panels, respectively. The unrelaxed cluster shown is a $M_{500} = 5.5\times10^{14}\,h^{-1}M_{\odot}$ system undergoing a near 1:1 merger, resulting in a large merger shock. Its mass has also increased by a factor of 5 since $z = 0.5$. In contrast, the relaxed cluster is a $M_{500} = 3.9\times10^{14}\,h^{-1}M_{\odot}$ system that has not experienced a major merger since $z = 0.6$ and has only grown by 9\% in mass since $z = 0.5$. The differences between the dynamical state of these clusters is also readily apparent in both the velocity and acceleration maps. The velocity field of the relaxed cluster is that of a quiescent system smoothly accreting matter from its environment. The vector field points predominantly toward the center of the system with larger velocities outside of $r_{500}$ and smaller velocities towards the core. In addition to the gas inflow, clockwise rotation of the ICM can be seen in the velocity field within $r_{500}$. On the other hand, the unrelaxed cluster has a much more varied velocity field. There is an outwardly propagating merger shock in the bottom left region of the map at about  $r\approx 1.4\,r_{500}$ with large associated velocities. As the shock passes through the ICM, it converts the bulk of the kinetic energy of the gas into thermal energy, decreasing the magnitude of the inward flowing gas velocity. This can clearly be seen in contrast between the small velocities of the gas in front of the shock and the large outward flowing velocities within the shock. This results in localized net outward acceleration seen in the acceleration field and hence a negative $M_{\rm accel}$ (Equation~\ref{eqn:jeans_accel}). A second shock can be seen in this system at the top right within $r_{500}$. This shock is not propagating in the plane of the map and therefore appears to have less ordered acceleration vectors in this slice. In addition to the two shocks, the cluster has large gas accelerations throughout $r_{500}$ induced by the ongoing merger. The corners of the map with very small acceleration vectors show regions in the outskirts of the cluster thus far untouched by the merger. Conversely, the acceleration field of the relaxed cluster has very small magnitudes to the point of being almost non-existent in the core of the system. While the velocities in both systems have comparable magnitudes, the acceleration vector fields paint very different pictures for the two clusters, suggesting that there exists gas dynamical information that cannot be probed with kinematic measurements alone. It is this additional gas dynamical information that we will characterize below.

	The top row of Figure~\ref{fig:mcompare} shows the hydrostatic mass bias $M_{\rm therm}/M_{\rm true}$ for mass enclosed within three radii, $r_{2500}$, $r_{500}$, and $r_{200}$.  A summary of the values of the mean bias and error on the mean for all mass estimates can be found in Table~\ref{tab:mbias}.  For the total sample of clusters, the hydrostatic mass bias varies from $-7\%$ to $-13\%$ over all radii with scatter varying slightly with radius.    

	We also look at the dependence of the hydrostatic mass bias on cluster dynamical state. Relaxed clusters are represented by blue circles and unrelaxed clusters are represented by red triangles in Figure~\ref{fig:mcompare}. For the relaxed clusters, the mean hydrostatic mass bias ($\pm 1\sigma$ error) becomes more negative with increasing radius, from $-5\%\pm4\%$ at $r_{2500}$ to $-11\%\pm4\%$ at $r_{200}$. On the other hand, unrelaxed clusters shows less negative mean hydrostatic mass bias going from smaller to larger radius: from $-23\%$ at $r_{2500}$ to close to zero at $r_{200}$, but the scatter increases from $\sim 14\%$ at $r_{2500}$ to $\sim 35\%$ at $r_{200}$. This shows that actively merging systems can have highly negative or positive mass bias depending on the location of the infalling subcluster or post-merger shock (N12). Moreover, these effects become more prominent at larger radii as the gas in the infalling subclusters has not undergone complete disruptions in the cluster outskirts, resulting in higher level of gas motions compared to the cluster cores.  Our measurements of the hydrostatic mass bias generally agree with previous theoretical works \citep[e.g.,][]{rasia_etal06,nagai_etal07b, jeltema_etal08, piffaretti_etal08, lau_etal09, meneghetti_etal10, nelson_etal12, rasia_etal12, suto_etal13}. 
		 
	Previous attempts to recover the true mass from hydrostatic equilibrium through the inclusion of gas motions have mostly assumed steady-state (i.e., $\partial {\bf v}/\partial t=0$). In the second row of Figure~\ref{fig:mcompare} we show the average reconstructed cluster mass neglecting the acceleration term, $M_{\rm accel}$. The steady-state mass reconstruction greatly reduces the magnitude of the mean hydrostatic mass bias for the full sample at all radii, consistent with previous work \citep{rasia_etal04, fang_etal09, lau_etal09, nelson_etal12}. For example, the mean mass bias at $r_{500}$ is reduced from $-10\%$ to nearly zero. However, this reconstruction does not reduce the scatter, and in fact it increases slightly from $17\%$ to $19\%$ at $r_{500}$.  Similarly, for the relaxed clusters, the magnitude of the mean bias is significantly reduced, but the scatter is slightly larger for masses at all radii. For the unrelaxed systems, the mean bias changes little at $r_{2500}$ from $-23\%$ to $-24\%$, but at $r_{500}$ and $r_{200}$ the mean bias increases from close to zero to $+13\%$. In other words, without gas acceleration the reconstructed mass {\em overestimates} the true mass at these outer two radii for the unrelaxed sample. The scatter is also systematically larger for the unrelaxed sample. Our results show that the mass reconstruction works well for reducing the mean bias but not the scatter for relaxed systems. Additionally the reconstructed mass overestimates the true mass in unrelaxed systems that have just undergone major mergers because of the presence of strong merger shocks (N12). 
	
	Next, we relax the assumption of steady-state by including the additional term $M_{\rm accel}$ due to gas acceleration to our mass reconstruction.  This acceleration term is negative for gas accelerating away from (or decelerating towards) the cluster center (cf. Equation~\ref{eqn:jeans_accel}). Shown in the bottom row of Figure~\ref{fig:mcompare}, the inclusion of the acceleration term has different effects on the mean and error of the hydrostatic mass bias at different radii for clusters in different dynamical states. For the complete sample, including the acceleration term decreases the magnitude of the mean bias at $r_{500}$ from $-10\%$ to $-4\%$, but leaves the bias unchanged at $-7\%$ at $r_{200}$.  Additionally, including the acceleration term decreases the scatter at all radii. For the relaxed sample, the inclusion of the acceleration term reduces both the mean bias and its $1\sigma$ error at $r_{2500}$ from $-5\%\pm 4\%$ to $-2\%\pm 1\%$. At $r_{500}$ ($r_{200}$), including the acceleration term reduces the mean mass bias to nearly zero and reduces its scatter from 7\% (17\%) to 4\% (9\%). For unrelaxed systems, including $M_{\rm accel}$ does not improve the mass reconstruction. In the inner regions, the acceleration term reduces the mean bias ($\pm1\sigma$ error) at $r_{2500}$ from $-24\%\pm 5\%$ to $-14\%\pm 4\%$, but makes the mean bias significantly more negative at $r_{500}$ and $r_{200}$. 
	
\begin{figure}[t]
\begin{center}
\includegraphics[scale=0.85]{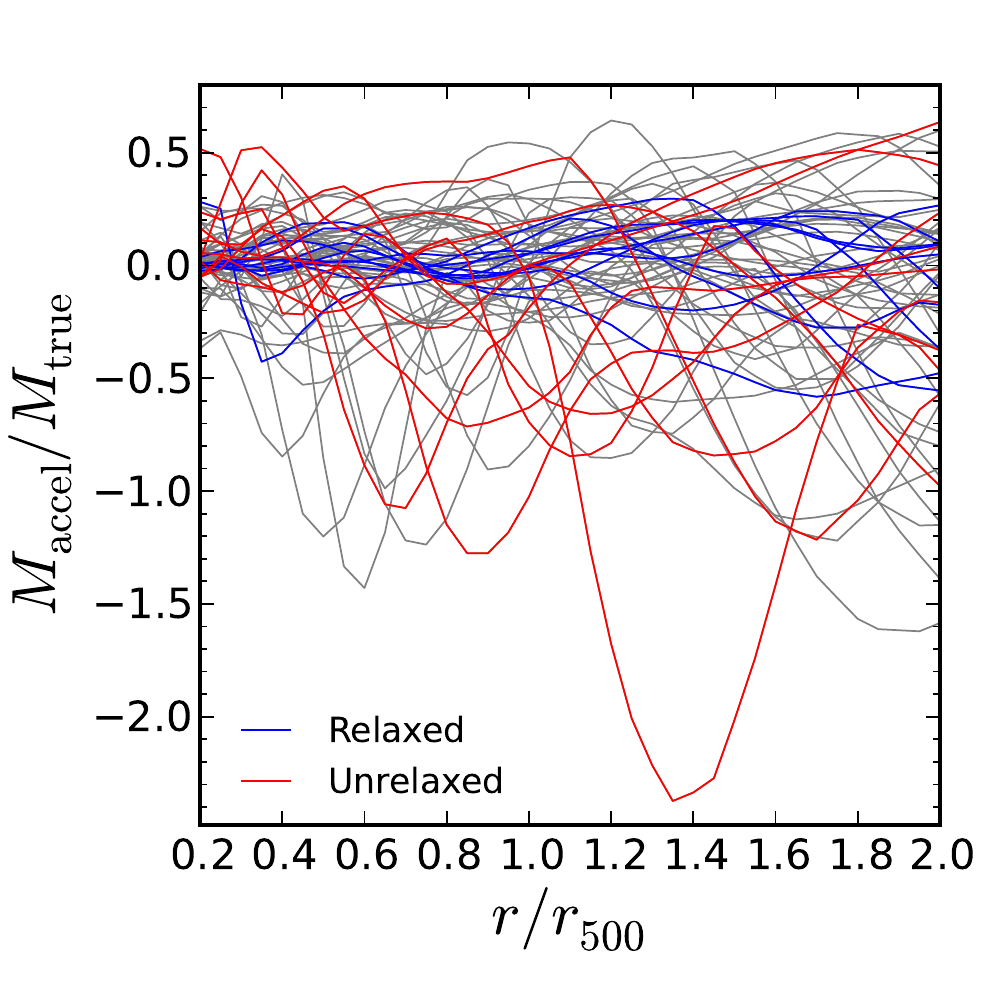}
\caption
{Comparison of $M_{\rm accel}/M_{\rm true}$ profile for relaxed (blue) and unrelaxed (red) clusters. The remainder of the sample is shown in grey.}
\label{fig:maccel_prof}
\end{center}
\end{figure}

	To characterize the nature of the acceleration term, we compare its fractional contribution $M_{\rm accel}/M_{\rm true}$ at three different radii in Figure \ref{fig:maccel}. We also plot the radial profile of $M_{\rm accel}$ in Figure \ref{fig:maccel_prof}. For the full sample, the mean contribution from $M_{\rm accel}$ increases with radii, from close to zero at $r_{2500}$ to $-10\%$ at $r_{200}$. Our results are consistent with \citet{suto_etal13} despite their smaller sample size and their different method in calculating $M_{\rm accel}$. For the relaxed systems, the mean $M_{\rm accel}/M_{\rm true}$  are consistent with zero at all radii. 

	The positive value of $M_{\rm accel}$ for the relaxed systems is consistent with ongoing smooth accretion -- accreted gas increases its velocity as it falls toward the cluster center. As the gas approaches the core of the cluster, the amount of acceleration decreases as infall is impeded by the ICM, and therefore $M_{\rm accel}$ approaches zero in the core of the cluster. Conversely, the unrelaxed clusters exhibit predominately negative $M_{\rm accel}$ values. The mean $M_{\rm accel}$  ($\pm 1\sigma$ error) becomes more negative with increasing radius: from $+10\%\pm 3\%$ at  $r_{2500}$ to $-25\%\pm 14\%$ at $r_{500}$ to $-39\%\pm 23\%$ at $r_{200}$. The large scatter in $M_{\rm accel}$ for the unrelaxed subsample is due to the varying positions and strengths of the troughs in the $M_{\rm accel}$ radial profile for different clusters, shown in Figure~\ref{fig:maccel_prof}. These localized troughs are driven by post-merger shocks, which cause sharp changes in the radial velocity. One such system is shown in Figure~\ref{fig:unrelaxed_example}, with a large shock at around 1.4 $r_{500}$.  The magnitude of $M_{\rm accel}$ depends on the strength of the post merger shock (which depends on the mass ratio, impact parameter and collision velocity of the merger).  As the shock propagates outwards during and after the merger, the trough in $M_{\rm accel}$ also propagates, resulting in the observed spread in $M_{\rm accel}$ for unrelaxed clusters at different merging stages. 

As $M_{\rm accel}$ becomes more negative with the strength of the merger shocks in the unrelaxed clusters, we expect it to compensate for most of the overestimates of the hydrostatic mass which also arises from the merger shocks. For systems in intermediate dynamical states, there are signs of such compensation which brings the total mass bias close to zero, particularly at $r_{200}$. This is not the case for the unrelaxed sample, however, possibly due to the the complicated geometries of the merging systems which can deviate significantly from spherical symmetry that we have assumed in our analysis.

Lastly, we investigate the redshift evolution of the hydrostatic mass bias and gas acceleration contribution. At higher redshift, mergers are expected to be more frequent and clusters to have enhanced mass accretion potentially leading to larger biases. In fact, we find that the hydrostatic mass bias and $M_{\rm accel}$ term have no trend with redshift within $r_{500}$.  However,  the biases at $r_{200}$ due to gas velocities and accelerations become more significant at higher redshift. At $z=0.6$ and $z=1.0$ the hydrostatic mass bias grows from $-8\%$ to $-14\%$ and $-19\%$, respectively. Clusters at higher redshift are more actively accreting, leading to significantly larger $M_{\rm accel}$ estimates of $-12\%$ and $-18\%$ at $z=0.6$ and $z=1.0$, respectively. However, the disturbed nature of the unrelaxed clusters leads to strong deviations from hydrostatic equilibrium so the mass recovery also becomes increasingly worse at earlier times from $-7\%$ to $-19\%$ at $z=0.6$ and $-33\%$ at $z=1.0$.

{
\def\m{\phantom{-}}
\begin{deluxetable*}{ccccccc}
\tablecaption{Biases in $M_{\rm therm}$, $M_{\rm tot} - M_{\rm accel}$ and $M_{\rm tot}$, and $M_{\rm accel}/M_{\rm true}$ measurements.\label{tab:mbias}}
\tablehead{
 & & \multicolumn{4}{c}{Mean~$\pm$ Error (1$\sigma$)\tablenotemark{a}} \\
\cline{3-6}\\[-1.7ex]
\colhead{Cluster mass} & 
\colhead{Sample\tablenotemark{b}} &
\colhead{~~$M_{\rm therm}/M_{\rm true}-1$ } &
\colhead{~~$(M_{\rm tot} - M_{\rm accel})/M_{\rm true}-1$} &
\colhead{~$M_{\rm tot}/M_{\rm true}-1$} &
\colhead{$M_{\rm accel}/M_{\rm true}$}
}
\startdata
 & all (62) & $-0.127\pm0.022$ & $-0.097\pm0.025$ & $-0.087\pm0.015$ & $0.010\pm0.031$\\
$<r_{2500}$ & relaxed (10) & $-0.050\pm0.035$ & $-0.034\pm0.037$ & $-0.015\pm0.007$ & $0.019\pm0.040$\\
 & unrelaxed (10) & $-0.225\pm0.045$ & $-0.237\pm0.057$ & $-0.142\pm0.041$ & $0.096\pm0.067$\\
\tableline\\[-1.0ex]
& all & $-0.099\pm0.021$ & $-0.012\pm0.026$ & $-0.044\pm0.019$ & $-0.032\pm0.035$\\
$<r_{500}$ & relaxed & $-0.092\pm0.021$ & $-0.014\pm0.023$ & $0.002\pm0.012$ & $0.016\pm0.030$\\
 & unrelaxed & $0.012\pm0.073$ & $0.128\pm0.092$ & $-0.121\pm0.083$ & $-0.248\pm0.144$\\
\tableline\\[-1.0ex]
& all & $-0.073\pm0.033$ & $0.029\pm0.042$ & $-0.067\pm0.027$ & $-0.095\pm0.056$\\
$<r_{200}$ & relaxed & $-0.110\pm0.042$ & $-0.042\pm0.055$ & $-0.017\pm0.029$ & $0.025\pm0.076$\\
 & unrelaxed & $0.011\pm0.116$ & $0.134\pm0.141$ & $-0.255\pm0.112$ & $-0.389\pm0.226$\\ [-2.0ex]
 
\enddata
\tablenotetext{a}{The scatter can be obtained from multiplying the error by $\sqrt{N - 1}$, where $N$ is the number of clusters.}
\tablenotetext{b}{The number of clusters in each sample is noted in the parentheses. The full sample is a mass-limited sample with a mass cut at  $M_{500} \geq 3\times 10^{14}h^{-1}M_{\odot}$ at $z=0$. The relaxed and unrelaxed subsamples are defined as the clusters having the lowest or highest 15\% fractional mass growth from $z=1$ to $z=0$ in the sample, respectively.}

\end{deluxetable*}

}

\section{Summary and Discussion}
\label{sec:summary}

In this work, we investigated the origin of the hydrostatic mass bias and the mass correction from gas motions using a mass-limited, cosmologically representative sample of 62 massive galaxy clusters from a high resolution hydrodynamical cosmological simulation. To date, the hydrostatic mass bias has been largely believed to arise from a non-zero gradient of non-thermal pressure provided by gas motions in galaxy clusters. We show that this is not the full physical description of the nature of the hydrostatic mass bias.  In addition to support due to turbulent and bulk gas velocities, the hydrostatic mass bias contains a non-negligible contribution from gas acceleration in the ICM in cluster outskirts with values depending on the dynamical state of the cluster. 

	We found that unrelaxed clusters, defined as having large recent mass accretion, exhibit significant bias due to gas acceleration, with magnitude comparable to the bias from non-thermal pressure associated with merger-induced turbulent and bulk gas velocities. Relaxed clusters, on the other hand, have small ($\lesssim 3$\%) acceleration bias, but the scatter in the mass bias can be reduced by accounting for gas acceleration.  Moreover, we found that the biases due to gas velocities and accelerations become more significant for high redshift clusters, where mergers are more frequent and clusters have more active mass accretion. Our work suggests the hydrostatic mass bias for individual clusters can only be corrected fully by accounting for both the gas velocities {\em and} accelerations.  Although for relaxed clusters the mean mass bias is consistent with zero for cases with and without the acceleration term, the error (and the scatter) on the mean bias is reduced in half with the inclusion of the acceleration term, suggesting that the mass recovery for individual relaxed systems can be improved by properly account for the contribution from gas acceleration. On the other hand for unrelaxed systems, the contribution from gas acceleration is significant, of order $\gtrsim 10\%$. Since in practice it is difficult to measure the gas acceleration directly through observations, the gas acceleration term introduces an {\em irreducible} bias in the hydrostatic mass estimates of galaxy clusters.

	Our work also suggests that the discrepancies between lensing and X-ray hydrostatic mass \citep[e.g.,][]{zhang_etal10, vonderLinden_etal12, mahdavi_etal13} cannot be fully explained without including gas acceleration and properly accounting for the dynamical states of clusters. This is especially true for high redshift clusters that are intrinsically less relaxed and hence have larger acceleration biases. However, their dynamical states are also difficult to measure owing to their general lack of photon counts in X-ray observations or lack of spatial resolution in current SZ observations. This introduce uncertainties in our definitions of ``relaxedness'' and the level of biases that must be accounted for in any given observational cluster sample. Furthermore, observed X-ray hydrostatic mass can also be biased low due to inhomogeneities in both gas temperature \citep{rasia_etal12} and gas density \citep{roncarelli_etal13}. These effects are more prominent at large radii, and they are also likely to be mass and redshift dependent. In order to improve the current cluster-based cosmological constraints, which are already limited by systematic uncertainties, all of these effects must be understood and characterized using detailed mock X-ray and lensing simulations.

	The hydrostatic mass bias can be partly accounted for by measuring gas velocities with upcoming high-spectral resolution X-ray observations or kinetic SZ measurements.  The mass recovery terms associated with gas velocities require full 3D gas velocity information.  The upcoming ASTRO-H Japanese-US X-ray satellite mission, scheduled to launch in 2015, will have a high-energy resolution calorimeter, capable of detecting the line-of-sight velocity dispersion from line-broadening measurements \citep[e.g.,][]{nagai_etal13}.  Although it is possible, in principle, to estimate tangential gas motions from resonant scattering \citep{zhuravleva_etal10}, full 3D velocity measurements will not be available in the near future.  

	The dynamical state classification scheme adopted in this paper, i.e., mass accretion history, is not unique, and is only applicable to simulated clusters. There are a number of observational probes of cluster dynamical states based on cluster morphologies such as power ratios and centroid shifts \citep[see, e.g., ][and references therein]{rasia13}. Line-of-sight velocity information from high-resolution X-ray spectra or kinetic SZ measurements may provide additional information in selecting relaxed clusters \citep[e.g.,][]{biffi_etal13}.  Additional work is required to characterize the connections between the observable dynamical state proxies and the mass accretion histories and the resulting effects on hydrostatic mass and cluster observables such as gas fraction.
		
	There are several caveats that must be kept in mind when interpreting our results. First, our simulations do not include physical viscosity, which might reduce the level of gas motions by dissipating the gas kinetic energy into the thermal energy of the ICM. If the physical viscosity is abnormally high, the gas would thermalize on a shorter timescale, thereby reducing the level of gas motions in the ICM. In the absence of physical viscosity the hydrostatic mass bias predicted in our simulated clusters can thus be interpreted as upper limits. We have also neglected radiative cooling, star formation and energy feedback from stars and active galactic nuclei. However, the exclusion of these extra physics is unlikely to have significant effects on gas accelerations, which are not important in the cluster core. Our simulations also do not include additional sources of non-thermal pressure from magnetic fields and cosmic rays. Although they are dynamically unimportant and only contribute to less than a few percent to hydrostatic mass bias, plasma effects like magnetothermal instability \citep[e.g.,][]{parrish_etal12} may drive non-negligible gas accelerations in relaxed clusters, particularly in the outskirts of clusters where there is a strong temperature gradient. Plasma effects could also change the thermal and dynamical properties of the ICM with time that could alter our predictions and interpretations of the acceleration bias.  For example, the acceleration term due to shocks can be sensitive to the effective dissipation scale of the ICM, which may be quite large in cluster outskirts if the shocks are collisionless. More investigation is needed to understand the plasma effects on the acceleration bias. Besides these physical processes, we have also assumed spherical symmetry in our analysis, which is not a good assumption especially for unrelaxed clusters. The deviation from spherical symmetry can affect the hydrostatic mass bias. It is also possible that the assumption of spherical symmetry is the cause of the poor mass recovery seen in the unrelaxed systems \citep[e.g.][]{chiu_molnar12,samsing_etal12}. Future work will address the effects of asphericity of cluster gas on the acceleration bias and investigate alternative methods of estimating the mass contribution from acceleration in disrupted clusters.
	
	Proper understanding of hydrostatic mass bias and its dependence of cluster dynamical state with redshift is crucial for cluster-based cosmological tests, as it directly affects interpretations of X-ray and SZ measurements and their cosmological inference.  Our work provides an additional step toward understanding the origin of hydrostatic mass biases and their impact on the use of galaxy clusters as one of the leading cosmological probes.

\acknowledgments 
This work was supported in part by NSF grants AST-1009811 and OCI-0904484, NASA ATP grant NNX11AE07G, NASA Chandra Theory grant GO213004B, the Research Corporation, and by the facilities and staff of the Yale University Faculty of Arts and Sciences High Performance Computing Center. LY also acknowledges support from the Yale Science, Technology and Research Scholars (STARS) Program. 

\bibliography{ms}

\appendix

\section{Comparison of Relaxation Criteria}
\label{sec:appendix}

Herein, we compare two quantitative proxies of the dynamical state of the clusters: (1) the mass accretion history, and (2) time since the last major merger. In the first proxy, we quantify the dynamical state of each cluster by the change in mass from $z = 0.5$ to $z = 0$: $\Delta M_{500} \equiv M_{500}(z=0)/M_{500}(z=0.5)$ (see Section~\ref{dynamicalstate}). We define the relaxed and unrelaxed subsamples as clusters with the smallest and largest $\Delta M_{500}$, respectively. In the second proxy, we measure the time from the last major merger $t_{\mathrm{merger}}$ for given cluster at $z = 0$. We define the relaxed and unrelaxed subsamples as the clusters with the longest and shortest $t_{\mathrm{merger}}$, respectively (see N12 for more details). 

\begin{figure*}[htbp]
\begin{center}
\includegraphics[scale=0.65]{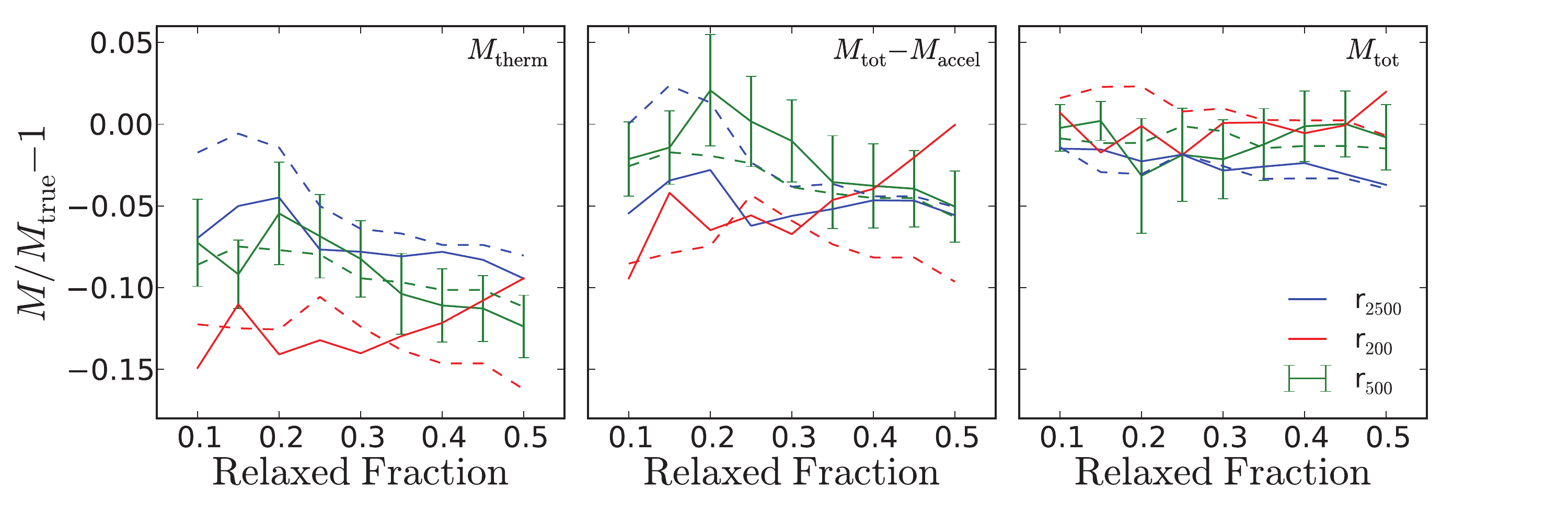}
\caption
{Comparison of $M_{\rm therm}$, $M_{\rm tot}$ - $M_{\rm accel}$ and $M_{\rm tot}$ (panels from {\em left} to {\em right})  for relaxed subsamples with varied percentage size of the full sample. The value of the respective mass estimates is shown for $M_{2500}$, $M_{500}$ and $M_{200}$ in blue, green and red lines, respectively. The relaxed fractions defined by the mass accretion history $\Delta M_{500}$ are shown solid lines, and those defined by the time since major merger $t_{\mathrm{merger}}$ are shown in dashed lines. Error bars denote the error on the mean for $r_{500}$, which are comparable to the errors for $r_{2500}$ and $r_{200}$. }
\label{fig:relax_cutoff}
\end{center}
\end{figure*}

 In Figure 6, we examine how the mean $M_{\rm therm}$, $M_{\rm tot}-M_{\rm accel}$ and $M_{\rm tot}$ (panel left to right) of the relaxed subsample depends on the definition of ``relaxed.'' Clusters are selected to be relaxed if they lie at the bottom $f_{\rm relax}$ of the distribution of $\Delta M_{500}$ (solid lines) for the mass accretion proxy, or if they lie at the top  $f_{\rm relax}$ of the distribution of $t_{\mathrm{merger}}$ (dashed lines) for the merger time proxy. We vary $f_{\rm relax}$ from 10\% to 50\%. The value of the respective mass estimates is shown for $M_{2500}$, $M_{500}$ and $M_{200}$ in blue, green and red lines, respectively.  The mass values are fairly insensitive to varying the relaxed fraction $f_{\rm relax}$ from 10\% to 30\% for both dynamical state proxies, with a small trend towards increasing bias with increasing $f_{\rm relax}$ as the subsample becomes contaminated with systems with intermediate dynamical states. At all radii, the mass bias is small for smaller relaxed fractions $f_{\rm relax}$ since by definition, these subsamples contain the most relaxed clusters and therefore the clusters that least deviate from hydrostatic equilibrium.  At higher $f_{\rm relax}$ the mass accretion subsamples include a number of intermediate dynamical state systems, which do not have fully relaxed gaseous atmosphere in the cluster outskirts. Specifically, for the $t_{\mathrm{merger}}$ proxy, at $f_{\rm relax}=30\%$  the relaxed samples contain clusters with $t_{\mathrm{merger}} < 4$~Gyr, which are shown to have significant hydrostatic mass bias (N12). The proxy based on mass accretion history is additionally sensitive to smooth accretion and minor mergers, hence it suffers from less contamination than the $t_{\mathrm{merger}}$ proxy. We note that the mass reconstruction $M_{\rm tot}$, which accounts for bias due to both gas velocities and acceleration, is able to recover the true mass to $\lesssim 5\%$ at all radii for $f_{\rm relax}<50\%$ regardless of dynamical state proxy. 

Through the comparison of dynamical state proxies and subsample definitions, we have determined that our results are generally insensitive to our choice of relaxation criteria. The minor instances of disagreement between the two methods, while not qualitatively affecting our results, highlight the different relative sensitivities to major mergers and smooth accretion/minor mergers.  Moreover, we have determined that for our work, the choice of 15\% lowest mass growth provides a statistically large sample while minimizing contamination of intermediate dynamical state systems in our subsample.  A more detailed examination of the relative effectiveness of accretion and merging events on sourcing non-thermal pressure as well as the characterization of various relaxation classifications, both theoretical and observational, will be explored in future work. 

\end{document}